\documentclass[a4paper,11pt]{article}
\pdfoutput=1 

\usepackage{jinstpub}
\usepackage{upgreek}
\usepackage{hyperref}

\title{\boldmath Large-area Si(Li) detectors for X-ray spectrometry and particle tracking in the GAPS experiment}

\author[a,1]{F.~Rogers,\note{Corresponding author.}}
\author[a]{M.~Xiao,}
\author[a]{K.~M.~Perez,}
\author[b]{S. Boggs,}
\author[a]{T.~Erjavec,}
\author[c]{L.~Fabris,}
\author[d]{H.~Fuke,}
\author[e]{C.~J.~Hailey,}
\author[d]{M.~Kozai,}
\author[b]{A.~Lowell,}
\author[e]{N.~Madden,}
\author[f,g]{M.~Manghisoni,}
\author[h]{S.~McBride,}
\author[f,g]{V.~Re,}
\author[f,g]{E.~Riceputi,}
\author[e]{N.~Saffold,}
\author[i]{Y.~Shimizu}

\affiliation[a]{Massachusetts Institute of Technology, Cambridge, MA 02139, USA}
\affiliation[b]{University of California at San Diego, La Jolla, CA 92093, USA}
\affiliation[c]{Oak Ridge National Laboratory, Oak Ridge, TN 37831, USA}
\affiliation[d]{Institute of Space and Astronautical Science, Japan Aerospace Exploration Agency (ISAS/JAXA), Sagamihara, Kanagawa 252-5210, Japan}
\affiliation[e]{Columbia University, New York, NY 10027, USA}
\affiliation[f]{Universit\`{a} di Bergamo, I-24044 Dalmine (BG) Italy}
\affiliation[g]{INFN, Sezione di Pavia, I-27100 Pavia, Italy}
\affiliation[h]{Space Sciences Laboratory, University of California at Berkeley, Berkeley, CA 94720, USA}
\affiliation[i]{Kanagawa University, Yokohama, Kanagawa 221-8686, Japan}

\emailAdd{frrogers@mit.edu}

\abstract{The first lithium-drifted silicon (Si(Li)) detectors to satisfy the unique geometric, performance, and cost requirements of the General Antiparticle Spectrometer (GAPS) experiment have been produced by Shimadzu Corporation. 
The GAPS Si(Li) detectors will form the first large-area, relatively high-temperature Si(Li) detector system with sensitivity to X-rays to operate at high altitude. 
These 10\,cm-diameter, 2.5\,mm-thick, 4- or 8-strip detectors provide the active area, X-ray absorption efficiency, energy resolution, and particle tracking capability necessary for the GAPS exotic-atom particle identification technique. 
In this paper, the detector performance is validated on the bases of X-ray energy resolution and reconstruction of cosmic minimum ionizing particle (MIP) signals. 
We use the established noise model for semiconductor detectors to distinguish sources of noise due to the detector from those due to signal processing electronics. 
We demonstrate that detectors with either 4 strips or 8 strips can provide the required $\lesssim$4\,keV (FWHM) X-ray energy resolution at flight temperatures of $-35$ to $-45^{\circ}$C, given the proper choice of signal processing electronics. 
Approximately 1000 8-strip detectors will be used for the first GAPS Antarctic balloon flight, scheduled for late 2021.
}

\arxivnumber{1906.00054}

\begin{document}
\maketitle
\flushbottom

\section{Introduction}
\label{sec:intro}

The General Antiparticle Spectrometer (GAPS) balloon experiment is designed to detect low energy (kinetic energy < 0.25\,GeV/n) cosmic antinuclei that could be produced in the annihilation or decay of dark matter particles in the Galaxy~\cite{DonatoAntideuterons, DonatoAntiprotons, Duperray, AntideuteronReview}. By analyzing data from three one-month-long Antarctic balloon flights, GAPS will produce a precision antiproton spectrum, observe or set leading limits on the flux of  antideuterons, and search for cosmic antihelium~\cite{Mori,HaileyGAPS1, HaileyGAPS2, AntideuteronAramakiSensitivity, AntiprotonAramakiSensitivity}. An abundant cosmic-ray background together with relatively low signal rates necessitate a detection method with a large geometric acceptance, high rejection factor, and low energy threshold. To meet these challenges, GAPS exploits a novel exotic atom-based detection technique. The detector consists of ten 1.6$\times$1.6\,m$^2$ planes of lithium-drifted silicon (Si(Li)) detectors \cite{Pell, Lauber, Goulding, Spieler} stacked with 10\,cm vertical spacing, surrounded by a  time-of-flight (TOF) system made of plastic scintillators. 

In the GAPS particle detection scheme, a low-energy cosmic antinucleus first passes through the TOF system, which measures its velocity and energy deposition, and provides timing information to within 500\,ps. It then traverses the layers of Si(Li) detectors, experiencing $\text d E/\text d x$ energy loss until it is captured by an atomic nucleus within a Si(Li) detector or aluminum support, forming an exotic atom in an excited state. This exotic atom de-excites through auto-ionizing and radiative transitions, emitting X-rays with energies determined by the reduced mass of the nucleus-antinucleus system and the atomic number of the target material~\cite{KEKBeam}. The antinucleus then annihilates with the nucleus, emitting pions and protons whose multiplicity scales with the antinucleus mass. Together, the characteristic X-ray energies, annihilation-product multiplicity, stopping depth given incident velocity, and energy deposition signatures uniquely identify an antinucleus species. Meanwhile, the abundant protons and other non-antimatter cosmic-ray particles are rejected based on their lack of hadronic annihilation products and X-rays. 

A Si(Li) system with sufficient stopping depth, spatial resolution, and energy resolution, combined with a TOF providing <500\,ps timing resolution, is key to the success of this detection technique. The Si(Li) detector system overall must be thick enough to serve as a target and stop antinuclei up to 0.25\,GeV/n, but each detector must be thin enough to have a large escape fraction for the de-excitation X-rays in the range of 20--100\,keV; a system of 10 layers of 2.5\,mm-thick detectors meets this requirement. They must cover a large area and have spatial resolution sufficient to distinguish tracks from incident particles and exotic atom annihilation products; a system of 1440 10\,cm-diameter detectors segmented into 4 or 8 active strips provides both~\cite{AntideuteronAramakiSensitivity}. The detectors must provide energy resolution of FWHM $\lesssim$4\,keV in the 20--100\,keV range in order to discriminate between the characteristic X-rays from the de-excitation of antiprotonic and antideuteronic exotic atoms. Finally, it is impossible to fly a pressure vessel or cryostat large enough for the GAPS instrument within the weight constraints of a long-duration balloon mission, and power is also limited. Therefore, the detectors must be operable at a relatively low bias, at ambient flight pressure, and at the relatively high temperatures of $-35$ to $-45^{\circ}$C.

The GAPS Collaboration has previously demonstrated successful operation of Si(Li) detectors that meet these requirements  in the prototype GAPS (pGAPS) balloon flight~\cite{DoetinchemPGAPS, MognetPGAPS, FukePGAPS}. The 10\,cm-diameter, 2.5 and 4.2\,mm-thick, 8-strip detectors used on pGAPS were acquired from the now-defunct SEMIKON Detector GmbH and were the first large-area Si(Li) detectors able to achieve <4 keV energy resolution at temperatures as high as $-35^{\circ}$C~\cite{Semikon}. Though the SEMIKON detectors met the geometric and performance requirements for the full-scale GAPS experiment, they were prohibitively expensive to produce in the large numbers required for GAPS, and the fabrication method was lost when the company went out of business. Accordingly, a new fabrication method for low-cost, large-area Si(Li) detectors was developed. An in-house fabrication method for 5\,cm-diameter, 1.25\,mm-thick, single-strip detectors was established  and used extensively to validate different production techniques~\cite{PerezSiLi}. Meanwhile, the authors in collaboration with Shimadzu Corporation developed a scalable procedure to produce the flight-geometry detectors whose performance is detailed herein. 

This paper discusses the performance of the first GAPS flight-geometry detectors. An overview of the production process, including the baseline detector performance and yield, are described in~\cite{KozaiSiLi}. The aim of this work is to demonstrate that both the flight-geometry 8-strip design and the alternate 4-strip design meet the requirements in terms of X-ray energy resolution and particle tracking performance needed for a GAPS flight. 
The details of the detector design and handling are discussed in sections~\ref{sec:design} and \ref{sec:prep}, respectively, while testing and performance is in section~\ref{sec:performance}. The setup for all tests is discussed in section~\ref{sec:setup}. In section~\ref{sec:bias} we show how capacitance 
measurements are used to set the $-250$\,V operating bias. 
The detector response to minimum ionizing particles (MIPs) is discussed in section~\ref{sec:tracking}. Energy resolution and calibration are assessed directly using radioactive sources, the results of which are detailed in section~\ref{sec:spectra}.

\section{Detector design}
\label{sec:design}

Details of the fabrication technique and process yield for the GAPS Si(Li) detectors produced at Shimadzu Corporation are described in~\cite{KozaiSiLi, KozaiIEEE}. The Shimadzu detectors differ from the SEMIKON ones in their ability to use silicon substrate from SUMCO Corporation rather than more costly substrate from Topsil Semiconductor Materials; their larger grooves machined using the easier and simpler technique of ultrasonic impact grinding rather than more costly plasma-etched grooves; their top-hat rather than inverted-T geometry, as defined in~\cite{Llacer}, which allows for simpler preparation of exposed surfaces; their readout from the \emph{n}$^+$-side rather than \emph{p}-side; and in the thin undrifted layer on the \emph{p}-side, which has proven critical to suppressing leakage currents in these large-area and high temperature detectors. 

\begin{figure}[htbp]
\centering 
\includegraphics[width=.76\textwidth,trim=100 20 130 280,clip]{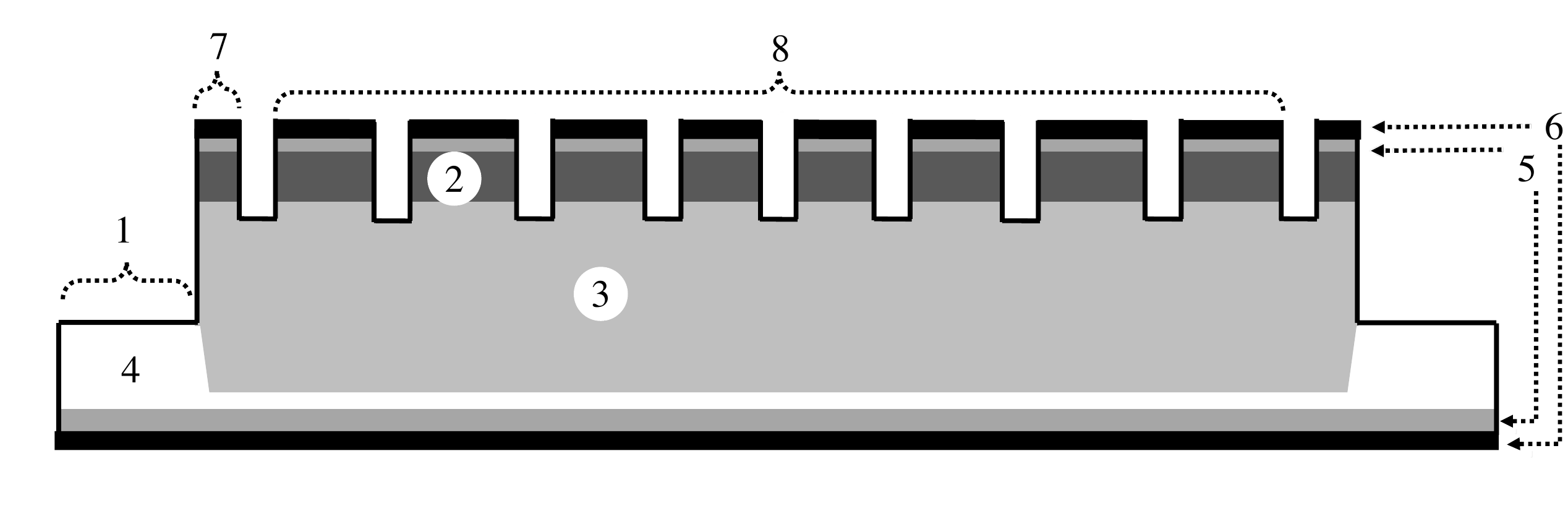}
\qquad
\includegraphics[width=.73\textwidth,trim=0 124 0 0,clip]{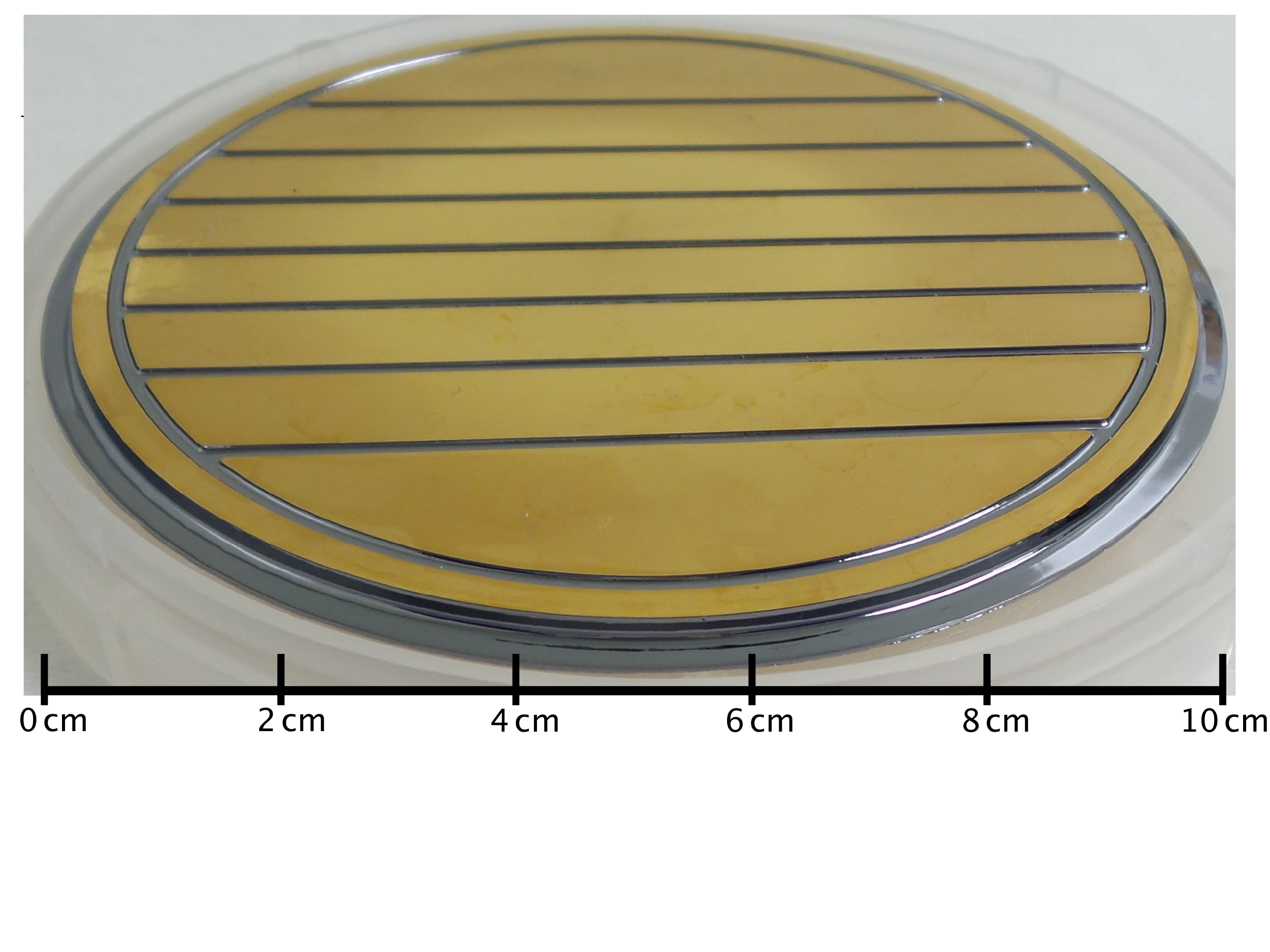}
\qquad
\caption{\label{fig:det} \emph{Top}: Diagram of the cross-section of an 8-strip GAPS detector (not to scale). The top-hat geometry is defined by removing Si from the top perimeter of the detector, leaving a $\sim$1\,mm-deep, $\sim$3\,mm-wide top-hat brim (1). 
Li ions from the $\sim$0.1\,mm-thick \emph{n}$^+$ Li-diffused layer (2) are drifted down through the \emph{p}-type wafer to form the compensated active volume (3). The Si in the top hat brim, and in a 0.1\,mm-thick region at the bottom of the detector, remains undrifted \emph{p}-type (4). The electrical contacts on top and bottom of the detector consist of a $\sim$20\,nm-thick Ni layer (5) topped with $\sim$100\,nm Au (6). The $\sim$1\,mm-wide, $\sim$0.3\,mm-deep grooves separate the guard ring (7) from the active region (8) and segment the round active region into parallel strips of equal area. 
\emph{Bottom}: Photograph of an 8-strip GAPS detector.
}
\end{figure}

Each Shimadzu detector begins as a $\sim$100\,mm-diameter, 2.5\,mm-thick wafer of single-crystal boron-doped \emph{p}-type silicon. Lithium is evaporated onto the top surface and thermally diffused through the material, forming an \emph{n}$^+$ layer. The top-hat geometry is defined by removing a $\sim$2\,mm-wide, 1.5\,mm-deep ring from the top perimeter of the detector, leaving behind a $\sim$1\,mm-thick region of undrifted \emph{p}-type material which we refer to as the "top-hat brim." Then, the evaporated and thermally diffused lithium is drifted through the bulk of the wafer, creating an active depth of $\sim$2.3\,mm of compensated drifted silicon sandwiched between the $\sim$0.1\,mm-deep layer of lithium-diffused \emph{n}$^+$ silicon on top and the $\sim$0.1\,mm-deep layer of undrifted \emph{p}-type silicon on the bottom. Both top and bottom of the detector are coated with $\sim$20\,nm of nickel and $\sim$100\,nm of gold, forming the electrical contacts. A circular 0.3\,mm deep, $\sim$1\,mm-wide groove machined into the top surface defines a $\sim$2\,mm-wide ring, known as the "guard ring," between the top hat brim and the $\sim$90\,mm-diameter active area of the detector. The guard ring geometry is a key element for the high-temperature operation of these large-area detectors. During operation, the bias is applied across the active region, while the guard ring is grounded. At biases $\gtrsim$ 100\,V, the electric field forms a depletion region along the groove between the guard ring and the active region, isolating the surface leakage current along the perimeter of the wafer, which can be many orders of magnitude larger than the bulk leakage current, from the detector readout~\cite{Llacer}. The active area is further divided into strips of equal area by a series of parallel $\sim$1\,mm-wide, 0.3\,mm-deep grooves. This geometry is illustrated in Figure~\ref{fig:det}. The dimensions reported here are for the 8-strip detectors discussed in this paper and reflect the final flight geometry; the 4-strip detectors were produced with smaller active areas of $\sim$86\,mm diameter.

Both 4- and 8-strip detector geometries have been developed. The 8-strip detectors have been chosen as the basis of the GAPS design as they offer several advantages. The smaller strip area results in smaller per-strip leakage current and capacitance, two of the dominant components of the overall noise, as discussed in section~\ref{sec:model}. With their smaller per-strip capacitance and other noise parameters consistent with those outlined in this paper, 8-strip detectors with per-strip leakage current <5\,nA can meet the GAPS energy resolution requirement using a custom ASIC, currently under development, for pulse shaping and detector readout. This ASIC requires less power than a discrete-component design, which reduces demands on the cooling system and allows the detectors to operate at lower temperatures, further reducing their leakage current and improving energy resolution. Additionally, the 8-strip design is characterized by better particle tracking performance, as the smaller strip size provides finer spatial resolution and the compatibility with ASIC readout minimizes the amount of inactive material in the tracker. 
An alternate 4-strip detector design that meets the GAPS experiment requirements has also been validated. Though the larger per-strip capacitance makes ASIC readout unfeasible for 4-strip detectors, the smaller number of strips permits the use of a discrete-component preamplifier, which is technically easier to prepare for flight but requires more power per strip. The preamplifier readout can reach the required energy resolution but with the cost of increased instrument heat load and inactive material that could distort particle tracks.


\section{Detector surface preparation, handling, and storage}
\label{sec:prep}

Si(Li) detectors are sensitive to environmental conditions, and care must be taken to prevent damage or degradation during handling and storage. A surface passivation process has been developed and validated that will be applied to all GAPS flight detectors after production at Shimadzu Corporation. Many of the prototype detectors that we report on here, however, have no surface passivation. We discuss below the storage, cleaning, and handling procedures used with these un-passivated detectors. 

The Si(Li) detectors produced by Shimadzu Corporation have a large area of exposed silicon in the grooves and top-hat brim (see Figure~\ref{fig:det}). Changes in the silicon surface state can occur due to exposure to humidity or organic contaminants. This can increase conductivity along the surface, increasing surface leakage currents and thus degrading the X-ray energy resolution, as discussed in section~\ref{sec:model}. In addition, dust or particulate contaminants on the bare silicon can change the electric field configuration along the groove, possibly increasing leakage current and affecting charge collection efficiency or cross talk. 
To mitigate damage due to these effects, the detectors are stored in a desiccant box with relative humidity maintained <10$\%$, and the laboratory space is maintained at <30$\%$ relative humidity. The detectors are handled only using clean wafer tweezers or gloves, and electronic components are chosen to be low-outgassing. Prior to testing, the exposed silicon surfaces are prepared by swabbing with ACS-grade methanol in a flowing nitrogen environment, which removes any particle or dust contamination and sets a light \emph{n}-type surface state. During the testing phase, the detectors are cooled under either vacuum or flowing nitrogen conditions to avoid condensation.

In addition to degradation by surface contamination, Si(Li) detectors are damaged by heat. The diffusion constant $D$ for lithium ions in silicon increases with temperature $T$ as
\begin{equation}
\label{eq:mobility}
D = 0.0023e^{-7700/T} [cm^2/s]
\end{equation}
as detailed in~\cite{Pell}. Thus the risk of harmful redistribution within the silicon lattice of the lithium ions in the \emph{n}$^+$ or compensated regions increases exponentially with temperature. Diffusion of lithium ions within a Si(Li) detector can damage a detector through two mechanisms. First, diffusion of the lithium ions from the \emph{n}$^+$ layer can increase the depth of the \emph{n}$^+$ region. If the \emph{n}$^+$ region spreads beyond the 0.3\,mm depth of the grooves separating the strips from each other and the guard ring, the strips and guard ring will no longer be electrically isolated. Second, movement of lithium ions in the silicon bulk can cause decompensation. The high resistivity of the bulk silicon is achieved by drifting the lithium ions through the \emph{p}-type silicon under a bias: during the drift, the lithium ions compensate the acceptor ions in the \emph{p}-type bulk as well as any inherent impurities in the silicon substrate. If the lithium ions diffuse away from these sites, the resulting decompensation can decrease the resistance of the bulk silicon, increasing the leakage current, or result in sites that trap electrons or holes, reducing the efficiency of charge collection. For long-term storage, the detectors will be arranged in airtight modules flushed with dry nitrogen gas and stored in a commercial freezer at $-25^{\circ}$C, where the diffusion constant for lithium is 150 times smaller than at room temperature.


\section{Performance of GAPS prototype large-area detectors}
\label{sec:performance}

\subsection{Experimental setup}
\label{sec:setup}

\begin{figure}[htbp]
\centering 
\includegraphics[width=.47\textwidth,trim=45 21 68 0,clip]{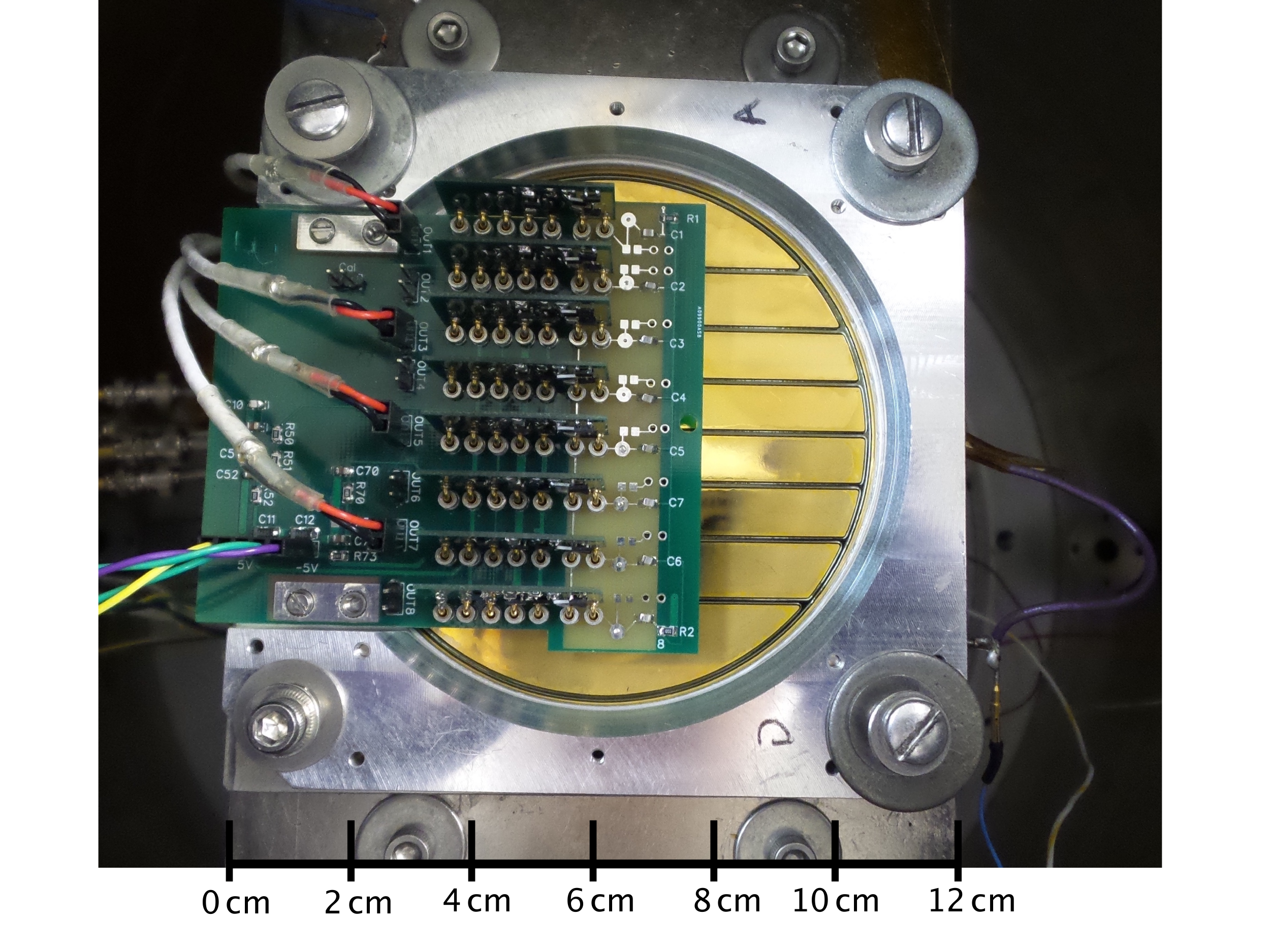}
\qquad
\includegraphics[width=.47\textwidth,trim=173 25 80 154,clip]{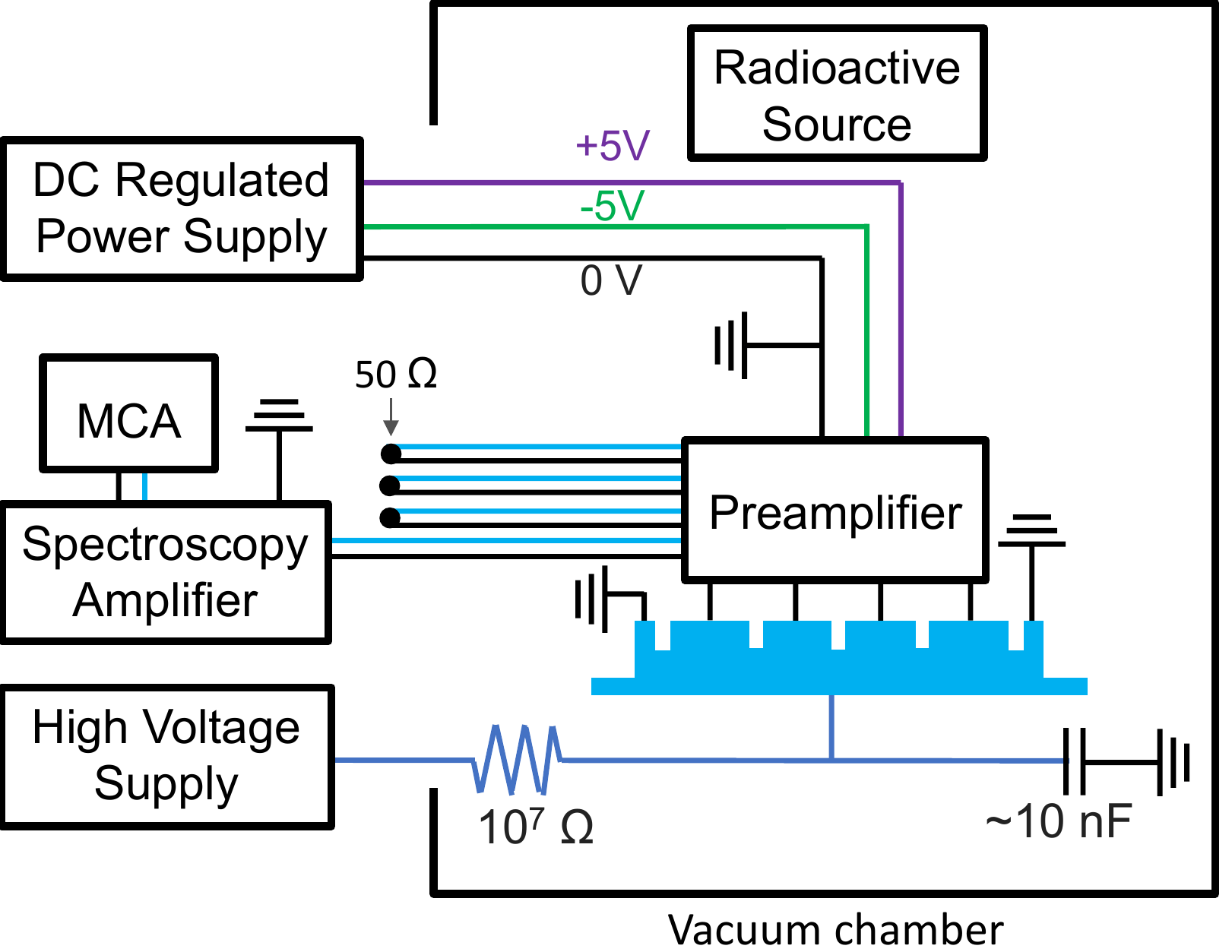}
\caption{\label{fig:mount} \emph{Left}: An 8-strip detector mounted in the setup for energy resolution measurements. \emph{Right}: The power and readout scheme, shown for a 4-strip detector. 
}
\end{figure}

Energy resolution measurements are performed in a custom aluminum vacuum chamber with pressure maintained below 2\,Pa using an oil-free scroll pump. The detector is held in an aluminum mount as shown in Figure~\ref{fig:mount}. This apparatus is bolted to a nickel-coated copper cold plate that is cooled by flowing cold gaseous nitrogen. Temperature is controlled by manually adjusting  the flow rate of nitrogen through the system and, with constant attention, can be stably maintained within $\pm$1--2$^{\circ}$C for time periods up to $\sim$30 minutes. Thermal and electrical contact to the detector is made by indium wire-mediated pressure between the detector guard ring and the mount, grounding the guard ring to mitigate the effect of surface currents on detector performance. A $\sim$ 5\,cm-tall aluminum cover placed over the preamplifier, detector, and mount acts as a Faraday cage, providing protection from electromagnetic interference pickup and any stray light. Two radioactive sources, 100\,$\upmu$Ci $^{241}$Am and 1\,mCi $^{109}$Cd in stainless steel housings, are used. The $^{241}$Am source rests on top of the aluminum cover $\sim$5\,cm from the surface of the detector. The $^{109}$Cd source is positioned $\sim$20\,cm from the detector on top of the vacuum chamber, so that the lower-energy lines of this higher-activity source are absorbed in the vacuum chamber material. The detector is biased from the \emph{p}-side at $-250$\,V by a Tennelec 953~HV supply fitted with an RC circuit that provides a local low impedance signal path and limits the maximum DC current as shown in Figure~\ref{fig:mount}. The negative bias voltage is supplied to the \emph{p}-side of the detector and isolated from the ground via a partially gold-plated ring of FR4. Temperature is monitored using a calibrated diode positioned on the detector mount and powered by a custom low-noise power supply.

The signal is read out from the \emph{n$^+$}-side by a custom 4- or 8-channel discrete-component charge-sensitive preamplifier board, which is pressure mounted to the strips via spring-loaded pins. Though a custom ASIC will ultimately be used for detector readout in final calibration and on the GAPS flights, a discrete preamplifier based on the architecture described in \cite{LorenzoPreamp} is used for detector testing while the ASIC is still under development. Each preamplifier channel consists of a 100\,M$\Omega$ feedback resistor, 0.5\,pF feedback capacitor, and a low-noise N-channel JFET with a capacitance of $\sim$10\,pF. The preamplifier is powered by $\pm$5\,V from a DC regulated power supply. The operating bias of +5\,V DC rail and the 100\,M$\Omega$ feedback resistor limit the per-strip leakage current to a maximum of 50\,nA before saturation. Signal from the preamplifier is processed by a Canberra 2020 Spectroscopy Amplifier with variable peaking time and digitized by an Ortec Ametek Easy MCA module. This system allows for readout of a single preamplifier channel; outputs for the remaining channels end in a 50\,$\Omega$ termination to prevent noise injection from external sources. A common ground from the NIM crate holding the spectroscopy amplifier is provided via the power supply to the preamplifier, the detector guard ring, and the RC circuit on the high voltage. This power and signal processing scheme is also illustrated in Figure~\ref{fig:mount}. 

For the cosmic muon spectral measurement, a slightly modified setup is used. The detector is cooled in a nitrogen atmosphere in an EC13 environmental chamber from SUN Electronics, allowing for automatic, stable temperature control. The spectroscopy amplifier and MCA are replaced by a CAEN N6725 digitizer, using 4\,$\upmu$s peaking time. The use of the digitizer allows multiple channels to be read out simultaneously and enables the use of coincident trigger conditions. All other power and readout components are the same as in Figure~\ref{fig:mount}.

Direct measurements of the capacitance and leakage current of each individual strip help determine the optimal operating bias and provide a comparison point for the noise model discussed in section~\ref{sec:model}. Strip capacitance is measured directly at ambient temperature, as capacitance at $-250$\,V varies only 1--2\% between room temperature and the typical operating temperature range of $-35$ to $-45^{\circ}$C. 
The guard ring and all unused strips are grounded while the capacitance of a single strip is measured using a HP 4280A meter, with bias supplied by an ORTEC 428 module. 
Per-strip leakage current is measured directly using a Keithley~487 picoammeter and voltage supply, with all other strips and the guard ring grounded.
For both capacitance and leakage current measurements, the preamplifier board is replaced with a direct connection to each electrode via pressure-mounted pins.


\subsection{Setting the operating bias}
\label{sec:bias}

An appropriate operating bias will be high enough to fully deplete the bulk of the detector, while minimizing both power requirements for this balloon-borne experiment and noise from leakage current, which increases with bias. A detector strip can be modeled as a parallel-plate capacitor with $C = \epsilon A/d$,where $\epsilon$ is the dielectric constant of silicon, 1.05$\times10^{-13}$\,F/m, $A$ is the strip area ($\sim$14.5\,cm$^2$ or $\sim$8\,cm$^2$ for a 4- or 8-strip detector, respectively), and $d$ is the depletion region depth. As bias increases, the depletion region grows, increasing the effective $d$ until the entire drifted depth is depleted and the capacitance approaches its asymptotic value, $\sim$72\,pF for a 4-strip detector or $\sim$35\,pF for an 8-strip detector. 

\begin{figure}[htbp]
\centering 
\includegraphics[width=.7\textwidth,trim=0 0 0 0,clip]{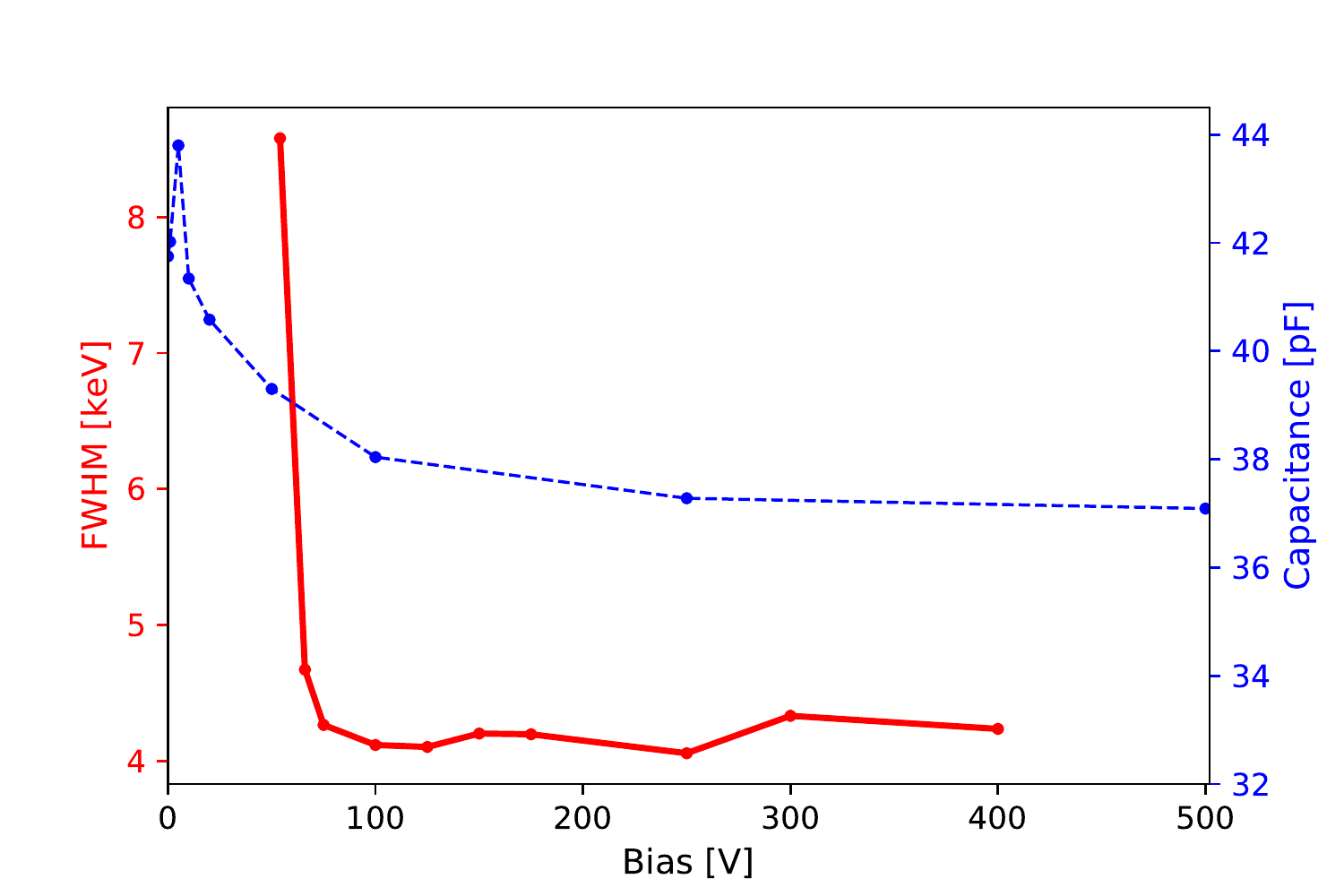}
\qquad
\caption{\label{fig:bias} Energy resolution (red solid) at 59.5\,keV as a function of applied bias, recorded using one strip of Sh0079 operated at -35$^{\circ}$C and processed with 10.8\,$\upmu$s peaking time, near the minimum of the resolution vs.\ peaking time curve for this high-capacitance setup. The energy resolution is affected as discussed in section~\ref{sec:model} by the capacitance (blue dashed) which decreases with increasing bias and the leakage current which increases with increasing bias. Based on the capacitance curve, the detector is fully depleted by $-250$\,V bias.
}
\end{figure}

Based on the capacitance measurements, the operating bias has been fixed at $-250$\,V. To validate this choice, the energy resolution at 59.5\,keV was recorded using a detector operating at a range of bias voltages from $-54$\,V, the lowest operable bias, to $-400$\,V. Figure~\ref{fig:bias} shows the energy resolution for a typical detector, along with the capacitance 
around the selected operating bias of $-250V$. Under these experimental conditions, the energy resolution is near minimum and the detector is fully depleted.

\subsection{Response to ionizing particles}
\label{sec:tracking}

The GAPS particle identification scheme relies on the Si(Li) detectors for tracking both incoming cosmic particles and outgoing annihilation products. In the laboratory, the Si(Li) detectors' tracking capability for charged particles is demonstrated using cosmic MIPs. A relativistic atmospheric muon vertically-incident on these 2.3\,mm active-depth Si(Li) detectors has a most probable value (MPV) of $\sim$750--800\,keV energy deposition from $\text d E/\text d x$ loss, while those arriving at greater angles deposit more energy.

\begin{figure}[htbp]
\centering 
\includegraphics[width=.7\textwidth,trim=0 0 0 0,clip]{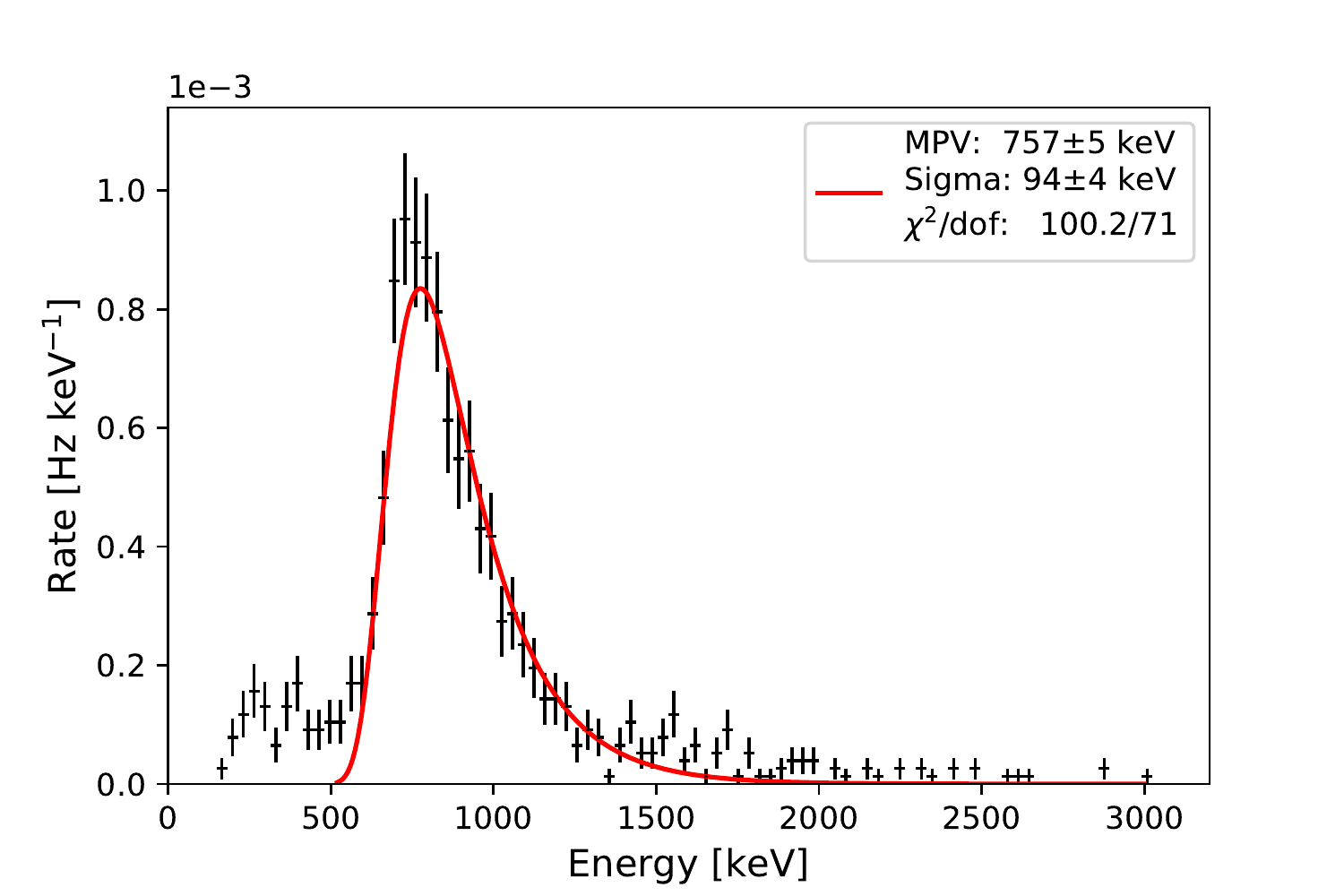}
\qquad
\caption{\label{fig:muon_spectrum} A spectrum of cosmic MIPs overlaid with a Laudau distribution fitted to the data. Details of the measurement and fit are given in section~\ref{sec:tracking}.
}
\end{figure}

The muon spectrum in Figure~\ref{fig:muon_spectrum} is produced by operating one strip of the 4-strip detector Sh0035 for $\sim$40\,min at a relatively high threshold of $\sim$200\,keV. To eliminate non-MIP background events and bias the sample toward vertical muons, a coincident signal is required with the corresponding strip of a second detector positioned $\sim$10\,cm below Sh0035. A Landau distribution, which describes fluctuations of energy deposition in the material, is fitted to the data, indicating a most probable value of $757\pm5$\,keV and a standard deviation of $94\pm4$\,keV. The calibration is performed based on the 59.5\,keV peak of $^{241}$Am and extrapolated to the higher-energy regime, introducing calibration uncertainty due to possible non-linear effects at higher energies. Still, the data in Figure~\ref{fig:muon_spectrum} are consistent with the expected distribution for atmospheric muons at sea level and they are within the required 10\% energy resolution for energy deposits of $1-100$\,MeV.

Antiprotons, antideuterons, and antihelium in the GAPS energy range are too slow to be MIPs and therefore will deposit more energy as they traverse the Si(Li) detectors. The different energy deposition signatures can be used for identification of the incident particle. To accomodate the different depositions expected from different particles as they slow to stop from up to 0.25\,GeV/n, the ASIC readout is designed to deliver energy deposition information in the range of 1-100\,MeV per strip with $\lesssim$10\% energy resolution. 

Cross talk due to electromagnetic coupling between the strips of a detector could reduce tracking or spectroscopy performance by splitting a signal from a charged particle between multiple strips or changing the amplitude of an observed signal. In a preliminary test using an anti-coincidence trigger between adjacent strips of an 8-strip detector irradiated by a $^{241}$Am source, energy resolution, peak location, and count rate at 59.5\,keV were consistent with and without the anti-coincidence requirement. However, detailed cross-talk studies of these detectors are ongoing, especially as pertains to the effect of cross-talk on charged particle reconstruction. We note that the per-strip count rate expected from the flux of cosmic ray particles and exotic atom annihilation products through the Si(Li) tracker is low relative to the $\upmu$s-scale readout time of the Si(Li) detectors. Therefore, cross talk is not anticipated to inhibit track reconstruction for GAPS.


\subsection{Spectral measurements} 
\label{sec:spectra}

\begin{figure}[htbp]
\centering 
\includegraphics[width=.7\textwidth,trim=0 2 0 31.4,clip]{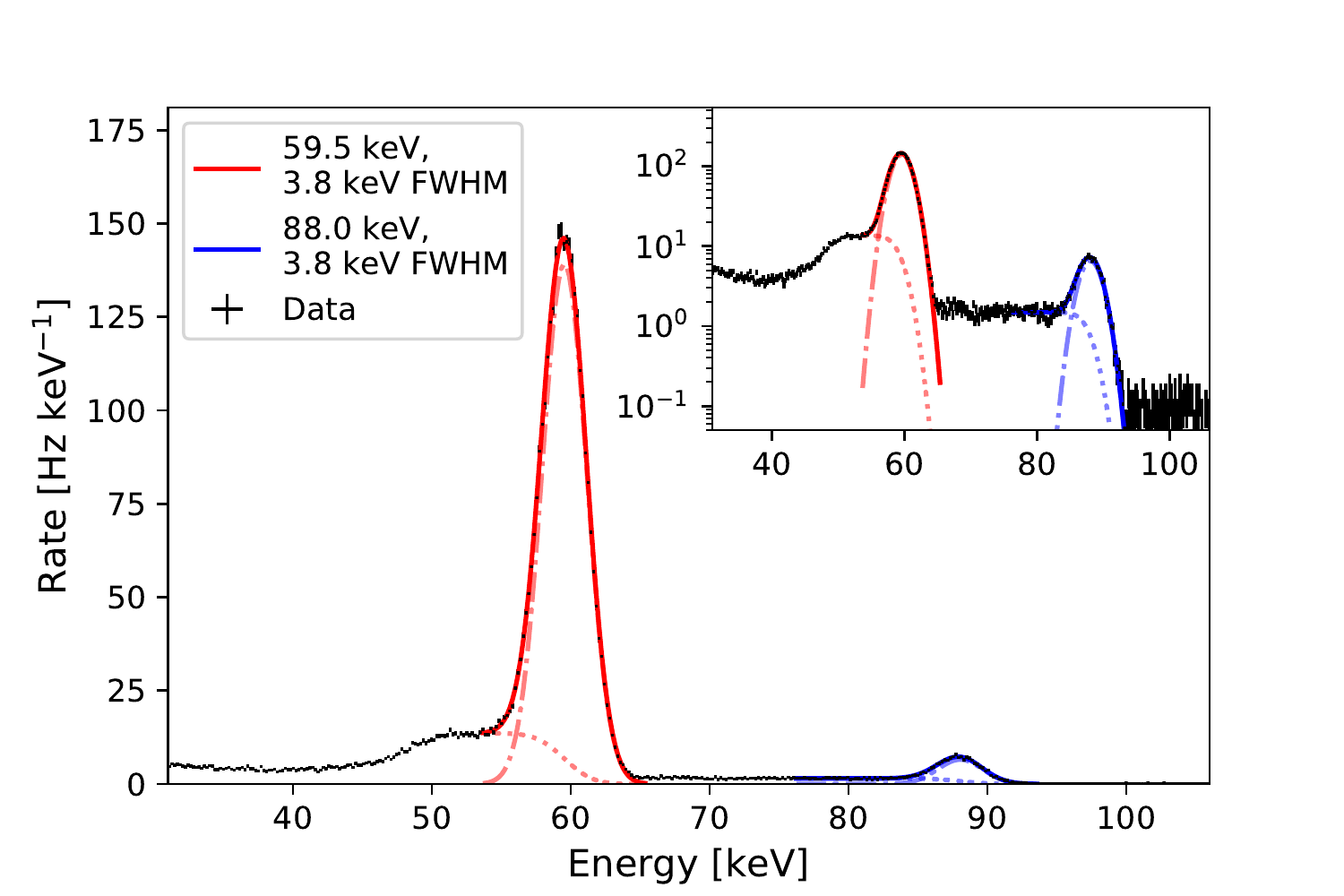}
\qquad
\caption{\label{fig:spectrum} Example spectrum of $^{241}$Am and $^{109}$Cd, recorded with one strip of Sh0025 at -35$^\circ$C and processed with a 4$\,\upmu$s peaking time. The data show each photopeak together with a low-energy tail of scattered $\gamma$-rays. The functional form is of a Gaussian (dash-dotted) plus an error function (dotted), as discussed in section~\ref{sec:spectra}. The inset shows the same data in semi-log format, to display the 88.0\,keV peak more clearly.
}
\end{figure}

The GAPS particle identification scheme relies on Si(Li) detectors with X-ray energy resolution of $\lesssim4$\,keV (FWHM) in the 20-100\,keV range to discriminate between the characteristic de-excitation X-rays of different antiparticle species. Figure~\ref{fig:spectrum} shows the response of one strip of the 4-strip detector Sh0025 to 59.5\,keV $\gamma$-rays from $^{241}$Am and 88.0\,keV $\gamma$-rays from $^{109}$Cd, demonstrating that the required energy resolution can be achieved at the relatively high temperature of -35$^{\circ}$C. Each photopeak is convolved with the Gaussian detector response. The low-energy feature to the left of each photopeak is due to Compton scattering from the surrounding materials. For a 59.5 or 88.0\,keV photon, the minimum scattered energy, corresponding to 180$^{\circ}$ backscatter, is 48.3\,keV or 65.5\,keV, respectively. 
 A function consisting of the sum of a Gaussian distribution to describe the photopeak, and an error function---convolved with the same energy resolution as the Gaussian---to approximate the nearly-flat higher-energy portion of the Compton scattering feature, is fitted to each peak of the spectrum. The range for each fit is  from the midpoint of the Compton scattering region to 6\,keV above the photopeak, or one FWHM above the photopeak if FWHM >6\,keV. The goodness of the fit is assessed using $\chi^2$ per degree of freedom as a figure of merit. 
Using the fitted position of the 59.5\,keV and the 88.0\,keV peaks, we find an offset from the zero energy intercept of $<2$\,keV.

\subsection{Energy resolution model}
\label{sec:model}

An energy resolution model allows us to disentangle the sources of noise due to intrinsic detector effects from those caused by the pulse shaping and readout electronics and to predict the performance of a particular detector under different conditions such as temperature. The energy resolution of a semiconductor detector read out via a charge-sensitive preamplifier and shaping amplifier is described by three terms. (1) A ``parallel noise'' term includes shot noise from the detector leakage current and thermal noise in any parallel resistance. This term typically dominates the noise at longer pulse peaking times. (2) A ``series noise'' term accounts for thermal noise from any series resistance and preamplifier FET noise. This term contributes the most noise at short pulse peaking time. (3) A ``$\frac{1}{f}$ noise'' term has equal intensity at all peaking times. The equivalent noise charge (ENC) that is read out is thus~\cite{Goulding, Spieler}:

\begin{subequations}
\label{eq:model}
\begin{align}
\label{eq:Emodel}
ENC^2& = \Big(2qI_{leak} + \frac{4kT}{R_p}\Big)\tau F_i + 4kT\Big(R_s+\frac{\Gamma}{g_m}\Big)\frac{C_{tot}^2}{\tau}F_{\nu} + A_fC_{tot}^2F_{\nu f}\,,
\end{align}
such that the FWHM energy resolution is given by:
\begin{equation}
\label{eq:FWHM_model}
FWHM = 2.35\epsilon\frac{ENC}{q} \,.
\end{equation}
\end{subequations}

In eqs.~\eqref{eq:Emodel} and \eqref{eq:FWHM_model}, $q$ 
is the fundamental electron charge, $k$ is Boltzmann's constant, $\epsilon$ is the ionization energy of silicon (3.6\,eV per electron-hole pair), and $T$ is the temperature, which we measure directly. $R_p$ is the parallel resistance of the preamplifier, 100\,M$\Omega$ in this setup, while $R_s$ is the sum of all series resistance with possible contributions from the preamplifier mounting method and the detector itself. The transconductance of the preamplifier input FET, $g_m$, is measured as 18\,mS at room temperature, and the constant $\Gamma$, related to the behavior of the channel in the JFET, is fixed to 1. Any small temperature variations in these parameters are absorbed into the complementary $R_s$ term for the purpose of fitting to our data. 
$A_f$ is the coefficient of $\frac{1}{f}$ noise, a temperature-dependent quantity that may include contributions from preamplifier noise, detector surface effects, or other electronic components. The total input capacitance, $C_{tot} = C_{det} + C_{FET} + C_{int}+ C_{stray}$, is the sum of all the parallel capacitance, including the individual strip capacitance $C_{det}$ (measured directly for each electrode), the capacitance of the preamplifier FET ($C_{FET}\approx10$\,pF), any inter-strip capacitance $C_{int}$, and any stray capacitance $C_{stray}$. $I_{leak}$ is the temperature-dependent leakage current of the strip. 
The dependence of each noise term on the particular pulse shaping system is parameterized by the form factors $F_i$, $F_{\nu}$, and $F_{\nu f}$. These are calculated as $F_i=0.367$,  $F_{\nu}=1.15$, and  $F_{\nu f}=3.287$ for our Sin$^4$ semi-Gaussian Canberra shaper, following~\cite{GouldingSig}, 
such that different components of the noise model can be evaluated by varying the peaking time of the spectroscopy amplifier. 

\begin{figure}[htbp]
\centering
\includegraphics[width=\textwidth,trim=86 520 86 84,clip]{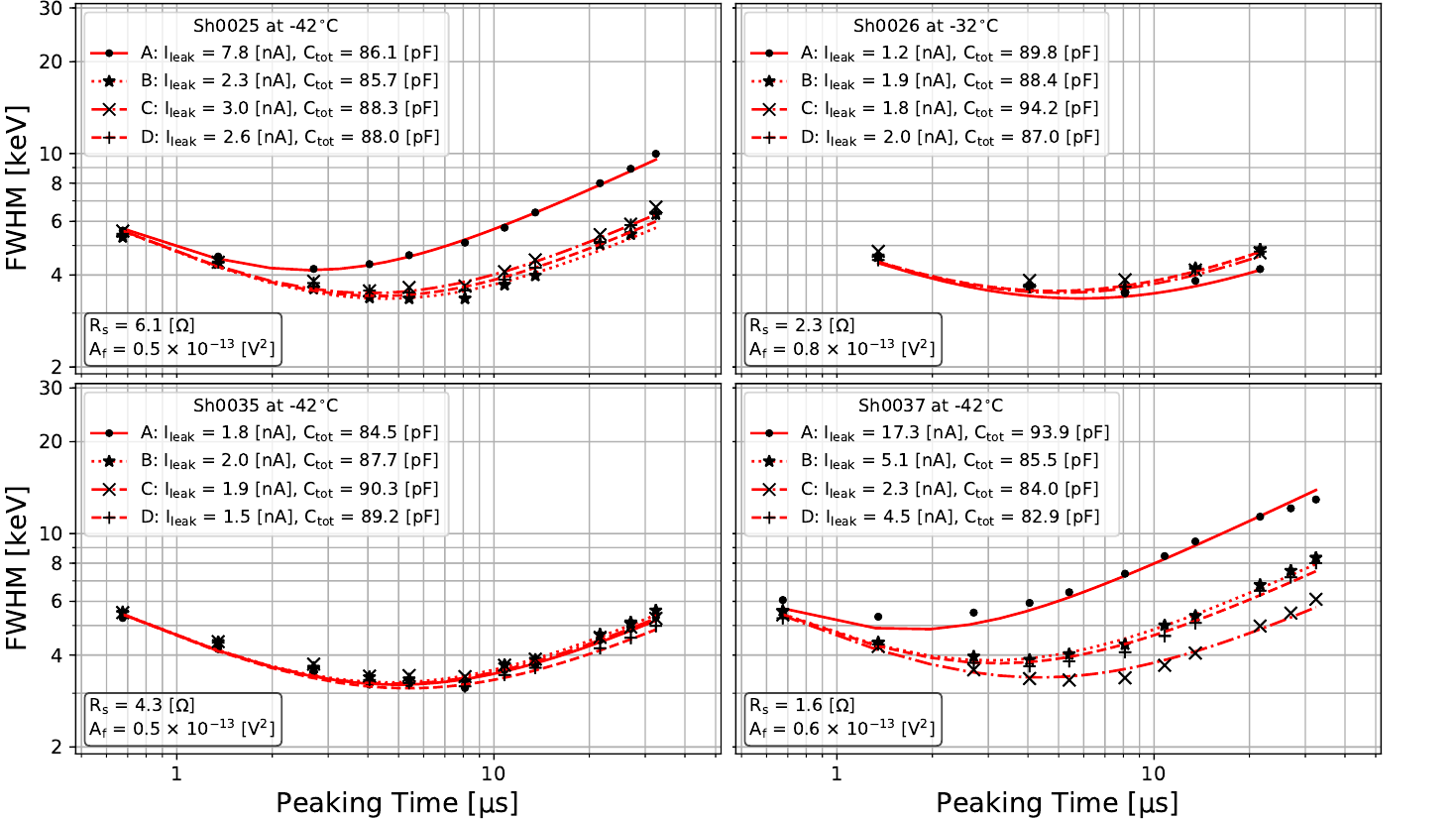}

\caption{\label{fig:modelstrip} Each panel above shows data for a single 4-strip detector, measured within or above the GAPS temperature range of  $-35$ to $-45^{\circ}$C. The measured energy resolution (FWHM) at 59.5\,keV is plotted as a function of peaking time for each strip A--D (black markers).  For each detector, the noise model (red lines, \eqref{eq:model}) can describe the data for all four strips, varying only $I_{leak}$ and $C_{tot}$ from strip to strip. $R_s$ and $A_f$ are fixed at their mean values from the fits for individual strips of each detector. The remaining noise model components are constant: $R_\mathrm{p}=100$\,M$\Omega$, $g_\mathrm{m}=18$\,mS, $\Gamma=1$, $F_\mathrm{i}=0.367$, $F_\mathrm{\nu}=1.15$, and $F_\mathrm{\nu f}=3.287$, as described in the text. }
\end{figure}

\begin{figure}[htbp]
\centering 
\includegraphics[width=\textwidth,trim=86 520 86 84,clip]{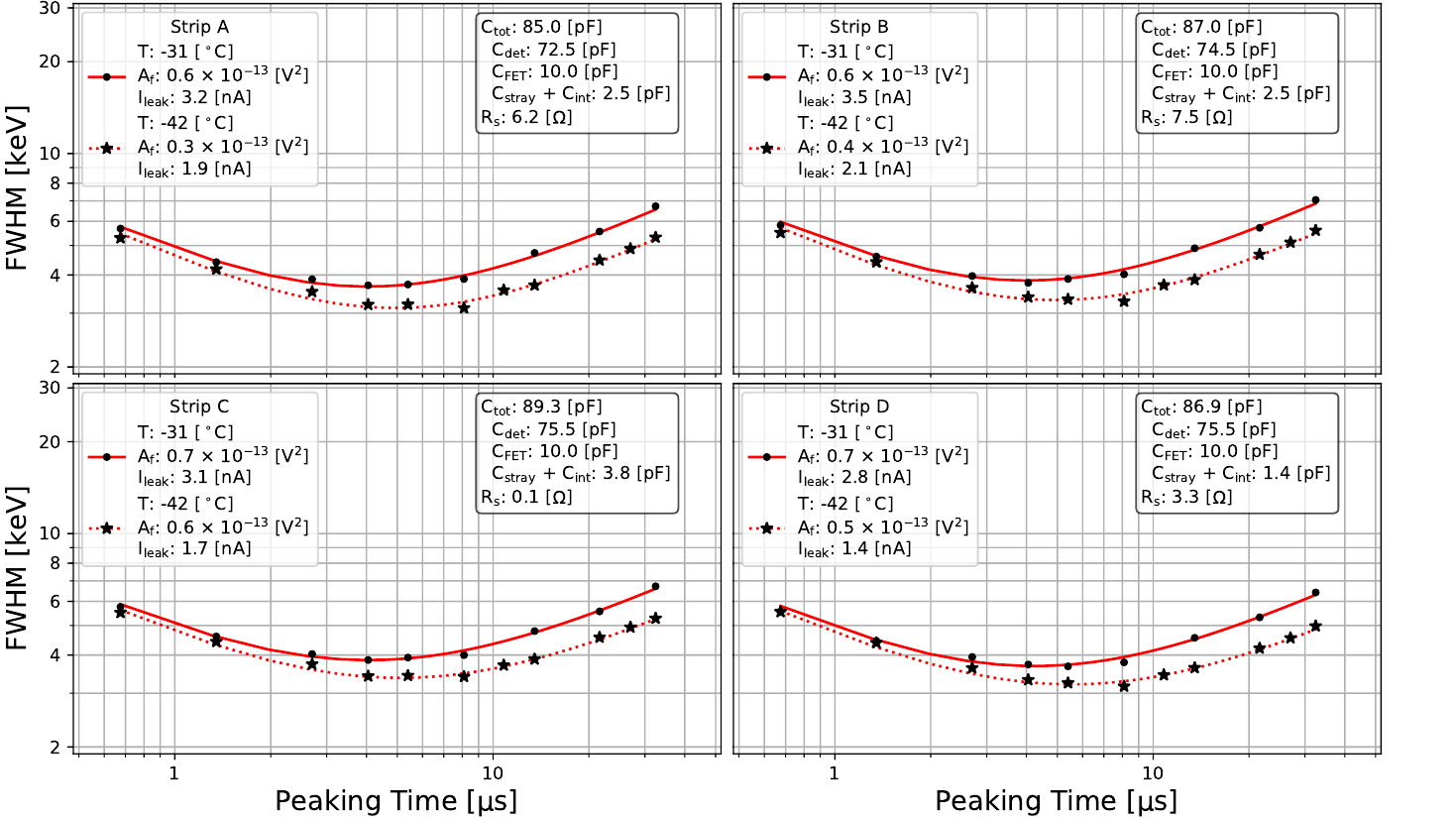}
\caption{\label{fig:modeltemp} Each panel shows data for one strip of the 4-strip detector Sh0035. For each strip, the measured energy resolution (FWHM) is plotted as a function of peaking time at two temperatures (black markers). The noise model (red lines, \eqref{eq:model}) can fit the data at both temperatures while changing only the parameters $A_\mathrm{f}$ and $I_{leak}$, which are expected to vary with temperature, in addition to $T$. The capacitance $C_{tot}$ and series resistance $R_s$ values are determined for each strip, and the remaining noise parameters are fixed as described in Figure~\ref{fig:modelstrip} and the text. }
\end{figure}

\begin{figure}[htbp]
\centering 
\includegraphics[width=\textwidth,trim=86 300 86 84,clip]{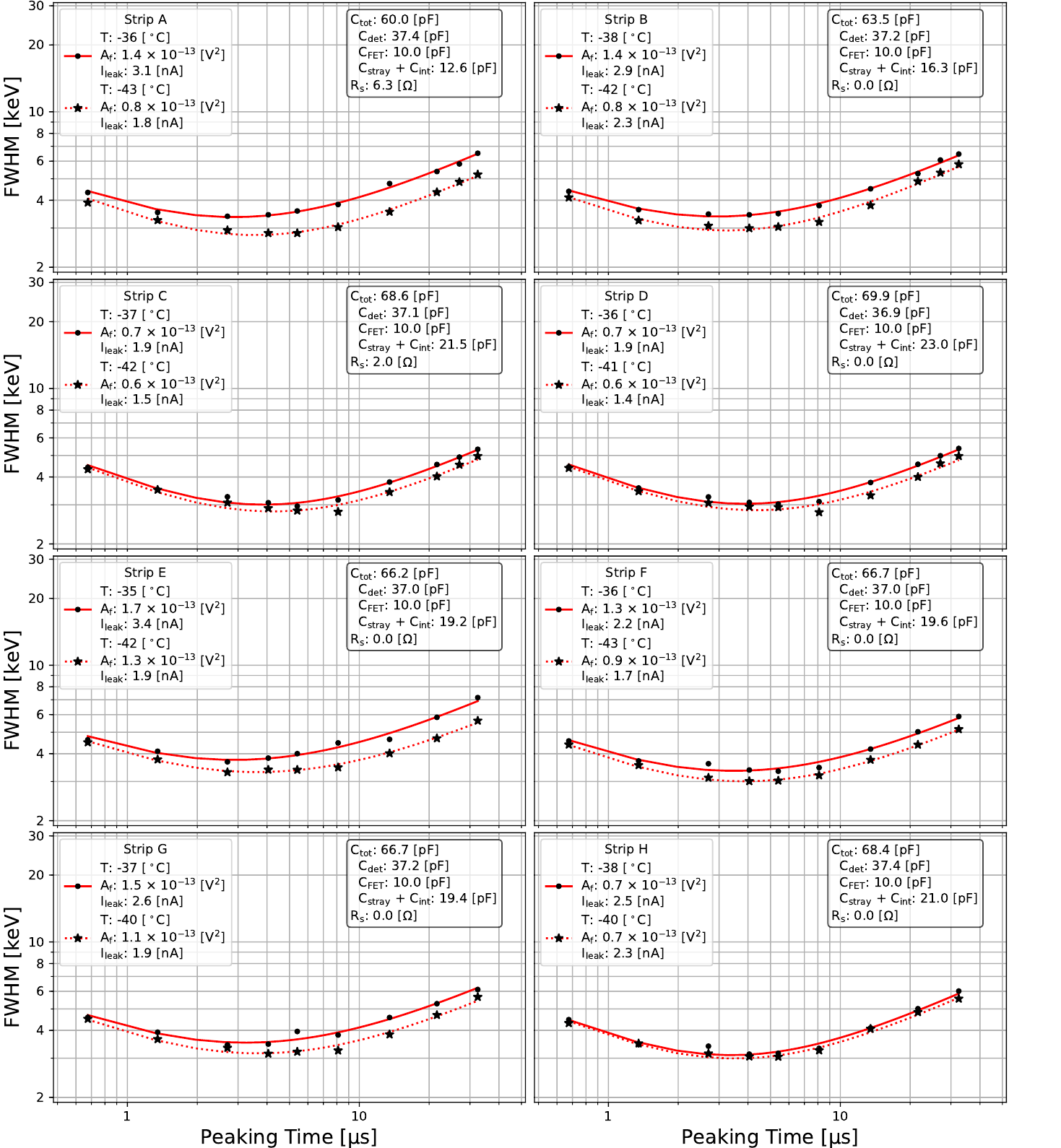}
\caption{\label{fig:modeltemp50} Each panel shows data for one strip of the 8-strip detector Sh0077. For each strip, the measured energy resolution (FWHM) at 59.5\,keV is plotted as a function of peaking time at two temperatures (black markers). The noise model (red lines, \eqref{eq:model}) can describe the data at both temperatures while keeping all parameters constant apart from $T$, $A_\mathrm{f}$, and $I_{leak}$, which are expected to vary with temperature. The capacitance $C_{tot}$ and series resistance $R_s$ values are determined for each strip independently while
the remaining noise parameters are fixed, as described in Figure~\ref{fig:modelstrip} and the text. The as-predicted temperature scaling indicates that based on calibration at only a few temperatures, we will understand detector performance at different temperatures during flight.
}
\end{figure}

The measured energy resolution as a function of peaking time is compared for all strips of four 4-strip detectors in Figure~\ref{fig:modelstrip}, for two operating temperatures of one 4-strip detector in Figure~\ref{fig:modeltemp}, and for two operating temperatures of one 8-strip detector in Figure~\ref{fig:modeltemp50}. 
These data are well-described by the energy resolution model by fitting the parameters $I_{leak}$, $A_f$, $C_{tot}$, and $R_s$ while keeping the others fixed at nominal values described above. 
To produce the plots shown in Figures~\ref{fig:modelstrip} -- \ref{fig:modeltemp50}, we first fit eq.~\eqref{eq:FWHM_model} to the energy resolution data for each strip at each temperature individually. Then, we combine the fitted values for each strip (Figure~\ref{fig:modelstrip}) or each temperature (Figures~\ref{fig:modeltemp} and \ref{fig:modeltemp50}) to produce the curves shown. For each strip at a given temperature, we derive the parameters $A_f$, $C_{int}+C_{stray}$, and $R_s$ from a fit to the measured energy resolution as a function of peaking time. Though $I_{leak}$ can be measured directly, we also fit this variable as a cross-check on the consistency of the fit. Because the first three parameters are degenerate, they cannot be fit simultaneously, so an iterative approach is used. First, since the $\frac{1}{f}$ component of the noise is constant in peaking time, it is fixed to a typical value of $0.6\times10^{-13}$\,V$^2$, while $C_{stray} + C_{int}$, $R_s$, and $I_{leak}$ are varied. 
Second, the value of $R_s$ is fixed at the best value from the first fit, and $A_f$ is instead varied. Finally, $C_{tot}$ is fixed at the best value from the second fit, and $A_f$ and $R_s$, along with $I_{leak}$ are varied. In each case, the previous best-fit values are used as the seed values for the next iteration. The $\chi^2$ per degree of freedom is used to assess goodness of fit. We confirm at the end of this fitting procedure that the best-fit leakage current is consistent with the directly measured value.

At a given temperature, the best-fit values for $R_s$ and $C_{tot}$ using the above procedure are consistent between all strips on a given detector to within a few percent as shown in Figures~\ref{fig:modelstrip} and \ref{fig:modeltemp}. This is as expected, since the preamplifiers are built to be identical, and the measured strip capacitance, $C_{det}\approx 73$\,pF, dominates $C_{tot}$ and is typically consistent to within $\sim$1\,pF between the equal-area strips, leaving $\sim$3\,pF stray and inter-electrode capacitance in addition to the 10\,pF from the FET. 

The $A_f$ parameter extracted from fits may contain contributions from multiple sources, including the discrete preamplifier and associated electronics, but is nonetheless consistently in the $0.5$ to $1.5\times 10^{-13}$\,V$^2$ range. Measurements of the preamplifier alone indicate that a large component of the total observed noise may be due to the readout electronics, which have not been optimized for low-noise operation; however, future measurements with a lower-noise preamplifier design are necessary to correctly identify all sources of $\frac{1}{f}$ noise.

Fixing $R_s$ and $A_f$ 
to the arithmetic mean of the best-fit values from the four strips, the energy resolution as a function of peaking time for each strip can be well described by eq.~\eqref{eq:model} by varying only the value of $I_{leak}$ and $C_{tot}$ for each strip, as shown for several 4-strip detectors in Figure~\ref{fig:modelstrip}. Thus, the parallel and series components of noise intrinsic to the detector can be clearly separated from those that depend on the readout, while the $A_f$ component can be attributed to a combination of the detector and the preamplifier.

For those 4-strip detectors in Figure~\ref{fig:modelstrip} with per-strip leakage current $<$10\,nA in the appropriate temperature region, the required energy resolution of $\lesssim$4\,keV FWHM was achieved. Two of the detectors (Sh0025 and Sh0037) have one strip with elevated leakage current. However, the remaining strips of those detectors reach the required energy resolution apparently free of deleterious effects due to their high-leakage current neighbor. 

The energy resolution at different temperatures can be consistently described using this model by varying only the temperature-dependent parameters $A_f$ and $I_{leak}$, as shown in Figures~\ref{fig:modeltemp} and \ref{fig:modeltemp50}. For each strip, the total capacitance $C_{tot}$ and the series resistance $R_s$, which show a weak dependence on the temperature, are fixed as the mean of the best-fit values at the two temperatures. $A_f$ and $I_{leak}$ are then fit, and $T$ is fixed to the measured temperature. 
For the 4-strip detector Sh0035, Figure~\ref{fig:modeltemp} demonstrates that the temperature variation of the energy resolution is well described by the noise model, varying only $A_f$ and $I_{leak}$.

The energy resolution at two temperatures for a flight-geometry 8-strip detector is shown in Figure~\ref{fig:modeltemp50}. The typical energy resolution for an 8-strip detector is improved relative to that of a 4-strip detector primarily due to the smaller strip capacitance and leakage current. The total capacitance of $\sim$60\,pF per strip ($
\sim$36\,pF detector capacitance plus $\sim$10\,pF FET and $\sim$14\,pF stray and interelectrode capacitance), reflects the reduction in area as compared with 86\,pF per strip for the 4-strip detectors ($\sim$73\,pF strip capacitance plus 10\,pF from the FET and $\sim$3\,pF stray). The additional stray capacitance for the 8-strip geometry is attributed to the geometry of the larger 8-channel preamplifier board used to readout the 8-strip detector positioned above the detector. Note that while the strip capacitance depends on the geometry, which is consistent between detectors of the same size and strip number, the leakage current can vary between strips and detectors, though in general $I_{leak}$ is expected to be lower for smaller strip area. All else being equal, the 8-strip design reduces the per-strip noise when compared to the 4-strip design, particularly at lower peaking times where capacitance drives the noise. Even with the noise contribution from the preamplifier or associated electronics, the energy resolution requirement $\lesssim$4\,keV FWHM was met. Development of a custom ASIC that will meet the energy resolution requirements given the detector characteristics described in this paper is underway~\cite{Manghisoni, Scotti}.

\section{Conclusions}
\label{sec:conclusions}

The first large-area, flight-geometry Si(Li) detectors that satisfy the unique performance requirements of the GAPS Antarctic balloon experiment have been developed in partnership with Shimadzu Corporation and validated by the GAPS collaboration. Their tracking performance has been validated using cosmic MIPs, and their energy resolution has been shown to meet the $\lesssim$4\,keV (FWHM) requirement in the energy range of 20-100\,keV and at the relatively high operating temperatures of $-35$ to $-45^{\circ}$C. We have demonstrated that the energy resolution as a function of peaking time and temperature follows the noise model of eq.~\eqref{eq:model}. The GAPS flight detectors will ultimately be read out by custom ASIC electronics currently under development, and the detector-ASIC combination will meet our energy resolution requirement given the detector performance achieved and reported in this paper. These Si(Li) detectors will form the first large-area, high-temperature silicon detector system with X-ray capability to operate at high altitude, and may also have additional applications, e.g., identification of heavy nuclei at rare isotope facilities~\cite{NSCL_frag,NSCL_Beta}. Production of >1000 8-strip detectors is ongoing for the initial GAPS flight, scheduled for late 2021, and since January 2019, we have been receiving flight detectors at a rate of $\sim$70 per month. 

\acknowledgments
We thank SUMCO Corporation and Shimadzu Corporation for their cooperation in detector development. We also thank the GAPS collaboration for their consultation and support. K.\ Perez receives support from the Heising-Simons Foundation and the Alfred P.\ Sloan Foundation. F.\ Rogers is supported through the National Science Foundation Graduate Research Fellowship under Grant No.\ 1122374. M.\ Kozai is supported by the JSPS KAKENHI under Grant No.\ JP17K14313. H.\ Fuke is supported by the JSPS KAKENHI under Grant Nos.\ JP2670715 and JP17H01136. M.\ Manghisoni, V.\ Re, and E.
\ Riceputi are supported by the Agenzia Spaziale Italiana (ASI). This work was partially supported by the NASA APRA program through Grant Nos.\ NNX17AB44G and NNX17AB46G. 

\bibliography{GAPS} 

\end{document}